# Floating Tip Nanolithography

**Alexander A Milner, Kaiyin Zhang and Yehiam Prior**

Department of Chemical Physics, Weizmann Institute of Science, Rehovot, Israel 76100

e-mail: alexander.milner@weizmann.ac.il

**Abstract.** We demonstrate *noncontact, high quality* surface modification with spatial resolution of ~20 nm. The nanowriting is based on the interaction between the surface and the tip of an Atomic force microscope illuminated by a focused laser beam and hovering 1-4 nanometers above the surface without touching it. The floating tip nanowriting is compared to mechanical surface scratching, and is found to be much more reproducible, and of higher quality. In an Apertureless Scanning Near Field Optical Microscope geometry the tip is illuminated by a focused femtosecond laser, leading to two different, clearly identifiable mechanisms for removing material from the surface: when heated by the laser beam, the hot-tip thermally patterns the surface of low melting temperature soft materials, and when focused right at the apex of the sharp tip, the enhanced electric field of the laser beam causes ablation in high melting temperature metal films.

Optical lithography has been the key enabling technology for microelectronics in the last few decades, but its spatial resolution is restricted by the optical diffraction limit to dimensions typical to the optical wavelength. Even the shortest UV wavelength currently in use (157 nm) does not provide the resolution required by the ever decreasing feature sizes of the "NANO" revolution. Reliable methods for sub-diffraction lithography are constantly being sought, and surface modifications by different interactions with the sharp tip of an Atomic Force Microscope (AFM) are prominent among them. In order of increasing complexity, the simplest of the three main methods used in this context is the direct mechanical scratching of a surface by a sharp tip, and the related technique of mechanically scratching of a thin over coating polymer layer followed by etching of the surface to be modified[1]. Next, the tip may be heated, resistively or by a focused laser beam[2,3] for a more pronounced effect on the surface. The use of hot tips was reported for thermal writing[4,5,6], for nano-indentation by resistively heated nano-size probes[7,8], or for the initialization of chemical transformations[9]. The third possibility in this hierarchy is based on the well known electromagnetic enhancement of a laser field in the vicinity of a sharp tip[10]. A different approach to surface modification is the high spatial resolution writing with foreign molecules (ink) on surfaces, as is done very well with the DPN (Dip Pen Nanolithography) method[11]. Note, however, that in DPN the surface itself is not modified, but rather material is added to it.

   The successful implementation of all these methods, as well as any other approach that will be based on a sharp AFM tip being in close proximity to the surface to be modified, require accurate control and knowledge of the Tip-Sample Gap (TSG). The introduction of Apertureless Scanning Near-field Optical Microscopy (ASNOM) with its obvious advantages of high spatial resolution based on the tip sharpness, and tolerance to high-density electromagnetic power channeled much of the current work towards commercial AFMs where tip-sample gap is controlled by monitoring the reflection of a laser beam from the back of the cantilever into a position sensitive photo detector[12,13,14]. While in principle such instruments should have provided a high degree of control over the tip position[15], in practice, the accurate determination of the TSG, especially in real time, is rather difficult. Furthermore, the task of harmlessly approaching a sharp tip to a solid surface and holding it at a constant distance of only a few nanometers is still quite impossible on standard AFMs. The conventional scheme consists of an approach to the surface till 'hard contact', followed by nominally raising the tip to the desired TSG. This approach does not work: with stiff cantilevers (~40 N/m) sharp tips are known not to survive the hard contact, and with soft tips (~2 N/m and less) the tip remains 'stuck' to the surface by capillary, electrostatic and other surface forces, and when released, it 'bounces' much farther than the desired few nm, typically to tens of nm above the surface (for further elaboration of this point, see Appendix A). As significant is the fact that the susceptibility of AFM piezo-ceramic scanners to temperature and humidity variations makes it practically impossible to maintain fixed TSG for a reasonable time without the constant verification by means of an active feedback loop. The non-contact operational mode of an AFM is rather useless for proximity tip surface interactions as it is based on a feedback loop that involves large amplitude oscillations at the mechanical resonance frequency of the cantilever, with a large mean distance between the tip and sample surface (typically 20-40 nm).

In this paper we demonstrate Floating Tip Nanolithography (FTN) - high quality non contact surface modification with high spatial resolution within a commercial AFM. Using the hot-tip effect, we write on a soft polymer (reliably, reproducibly and without physical contact between the tip and the surface) lines of ~20 nm width and 2-4 nm depth. When compared to the writing by mechanical scratching, our results clearly demonstrate the advantages of the newly introduced noncontact FTN methodology. We further use the same approach for noncontact writing on a gold film, where the physical effect leading to the material ablation is the electromagnetic enhancement of the femtosecond laser field under the sharp tip. In what follows, we show and discuss the advantages of the noncontact nanowriting, both by the hot tip and the electromagnetic enhancement effects. In both these sections, for the clarity of the presentation, we assume that we are able to control the TSG at will, and to position the tip at the desired gap above the surface. We postpone to the 'experimental methods' section the discussion of the newly introduced method enabling us to keep the tip hovering above the surface at a constant TSG of only a few nanometers for long durations of time and without ever touching the surface. The technical details of the method, as well as the calibration procedures we had to develop, are described in Appendix A.

**Hot tip lithography**

To realize FTN with a hot tip, we illuminated the tip with a beam of an external femtosecond laser. The laser light was focused in an ASNOM configuration by a 15 mm lens at an incident angle 70° to a small spot (diameter of 3-5 µm) a few microns above the tip apex so as to prevent any possibility of surface modification by direct interaction with the laser beam. The surface consisted of a silicon wafer over which a thin layer (a few microns) of a standard photoresist (AZ4620) was spin-coated. This polymer is known[16] to survive baking at 200°C, and we had to heat the tip well above this temperature to see any effect (see Appendix B). Experiments with a higher temperature compound, AP2210A (thermal decomposition temperature 518°C), under identical conditions showed similar results, but required much higher tip temperatures.

Figure 1a depicts a typical result of the floating tip nanolithography by a hot tip. Lines of ~ 20 widths and ~3 nm depth are shown, the writing is very clean, the material is clearly removed (ablated) away, and at the resolution afforded by these measurements there is no visible Heat Affected Zone (HAZ) around the ablated lines. Figure 1b presents the equivalent situation for lithography by mechanical scratching of the same surface. Based on our methodology, we can control the actual pressure applied by the tip on the surface (see below), and in order to make a valid comparison, we used the following procedure: the experiments were run several times at different (increasing) pressures, and the results examined. For low pressures, the scratching was barely visible, becoming progressively deeper, till at some point, the tip breaks. The results shown in figure 1b are the best that can be compared in terms of linewidth and depth to the ones in figure 1a. Under these conditions, after several repetitions of the experiment, the tip was worn out or damaged as was verified either by direct SEM observations of the tip or by the deterioration in resolution of the scanned images. The differences between the noncontact nanowriting and the best mechanical scratching are clearly seen in the line profiles (figures 1c and 1e): in the case of the mechanical scratching, the material removed from the line is piled up on the side. When the entire written area is profiled and digitally processed carefully, we find that under conditions of mechanical contact the material volume was preserved, as distinct from the FTN case where material is removed. This effect is demonstrated clearly at the intersection of two lines shown at larger magnification: whereas in the FTN writing (figure 1d), the two lines cross without disturbing each other, and from looking at the picture one cannot tell which line was written first; in the case of the mechanical scratching, (figure 1f), the second line overwrites the first and the 'removed' material covers the "trench" dug by the first, displaying a clear advantage for the hot tip noncontact nanolithography.

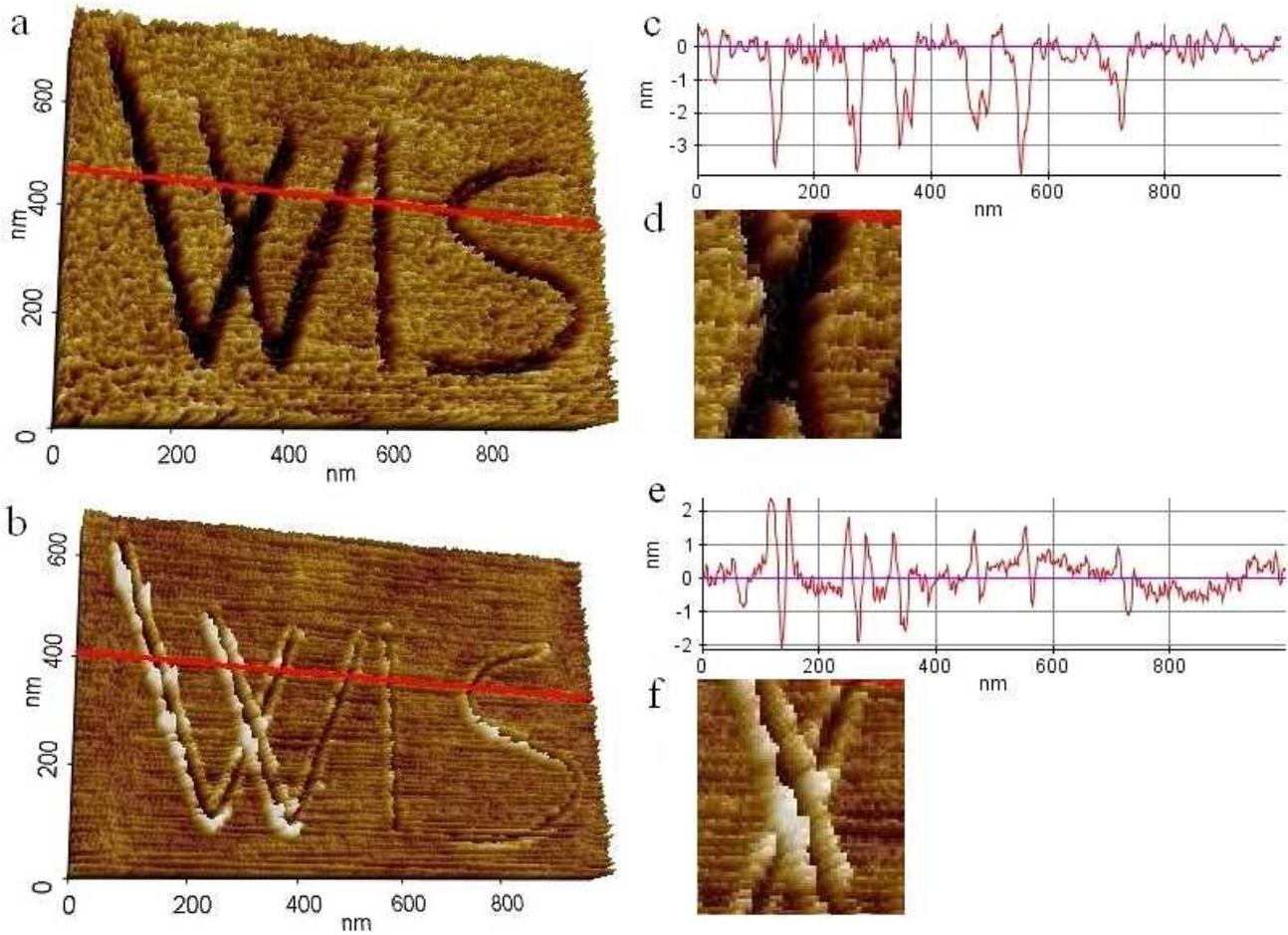

**Figure 1.** a) Hot floating tip writing WIS (for the Weizmann Institute of Science) on AZ4620 photoresist film. The writing speed was 50 nm/sec at tip-sample gap 3±2 nm and the average laser power was ~3x$10^5$ W/cm$^2$. b) Direct mechanical scratching of the same polymer with an identical tip. Profiles c and e correspond to the red cross-section lines in panels a and b; d and f are the magnified line intersections in a and b, respectively.

The letters written in figure 1 are hundreds of nanometers in dimension, and it is therefore difficult to appreciate the advantages of the noncontact FTN in terms of accurate lithography. In a second set of experiments, we opted to write much smaller shapes, in an attempt to check the limits of resolution for the method. In figure 2 we present the results of such experiments: a square, a circle and a pound sign were written, by FTN of a hot tip, and by direct mechanical scratching. The differences are even more pronounced than before. Whereas the FTN writing (figure 2a) maintains the desired shape, the shapes in figure 2b are distorted. Careful volume integration reveals again that no material was removed by the mechanical method, it was "pushed" according to the direction of motion of the tip, whereas significant amount of material was removed by the hot, floating tip. Moreover, while the shapes on the left are symmetrical, the shapes on the right show the effects of cantilever torsion and flexure which vary for different directions of relative movement.

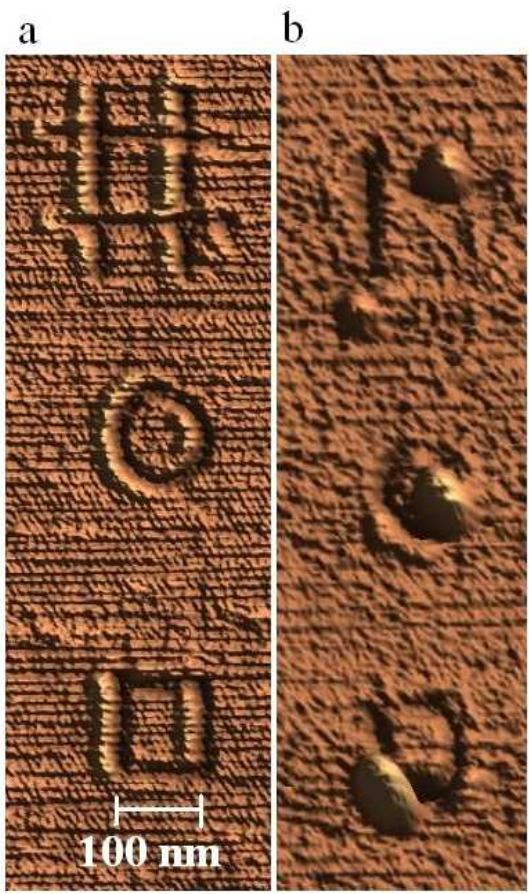
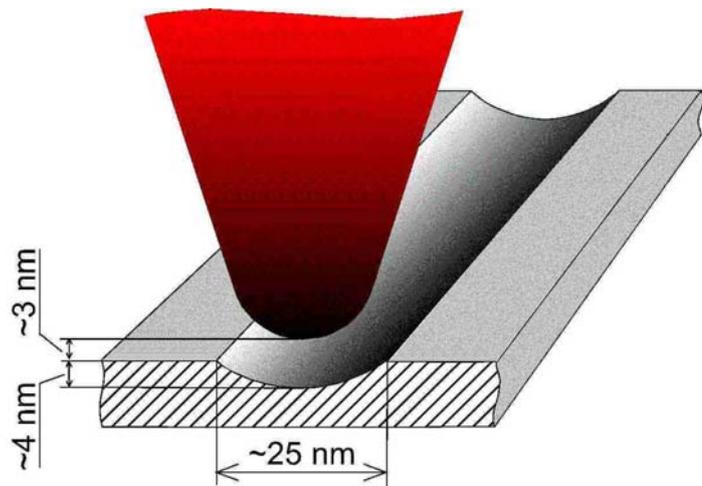

**Figure 3.** Idealized shape of the hot-floating-tip-produced trench with the parameters extracted from the experimental data presented in figure 1a.

**Figure 2.** AFM performed writing of small forms on the polymer surface: a) by means of laser-heated tip in the FTN regime and b) by direct mechanical scratching.

The schematic drawing in figure 3 provides an explanation for the operation of FTN with a hot tip. A sharp tip, with a radius of curvature of 10 nm (as per its specifications) is maintained at a tip sample gap of a few nm (i.e. 3nm in the figure). The tip is heated by a laser beam focused to a spot a few microns above the apex and the heat is efficiently transferred from the tip apex to the surface, causing the local surface temperature to rise well above the melting point of the polymer.
Several works have shown that under ambient conditions, at a separation of a few nanometers the heat flow from the hot tip to the surface is substantial[17,18,19]. Moreover, McCarthy et al.[20] measured the temperature of a tip illuminated by a focused laser beam, and have shown that even under very mild tip irradiation, the temperature rise may reach hundreds of degrees. Our own numerical simulations verify that for the actual conditions of our laser irradiation, the tip indeed reaches a temperature of a few hundred degrees, enough to affect the polymer surface (see Appendix B for details).
    The ability to use hot tips for surface nanostructuring was recognized a long time ago, and even considered for commercial applications in data storage in the IBM Millipede system[21]. However, contact between the tip and the surface often leads to undesirable results, hence the advantage of operating the AFM in a truly noncontact mode.

**Field enhanced lithography**

The next example we wish to consider is surface modification by the enhanced electric field under the sharp tip. Metallic surface structuring by an AFM tip illuminated by an ultrashort laser has been previously reported, and the literature contains two conflicting explanations for its mechanism. The first relies on material ablation due to plasmonic electromagnetic field enhancement under the tip[10], and the alternative explanation is based on rise in tip temperature due to laser absorption, and the ensuing tip elongation which causes the hot tip to mechanically hit the surface[2] and chisel away material. Both mechanisms are operative under most 'standard' conditions of irradiation of an ASNOM tip by a strong laser beam, and for the first time we can separate them by being absolutely sure the tip is never in mechanical contact with the surface.

The result for the modification of a gold surface is presented in figure 4. The tip was floating approximately 3 nm above the surface, laser light, polarized along the tip axis, was focused on the tip apex, and it was independently verified that the laser alone (without the tip) does not affect the surface. The tip temperature was much lower than the melting temperature of gold, so that the "hot-tip effect" can be safely disregarded. The observed written lines are ~20 nm wide and ~1 nm deep, material is actually removed and does not accumulate near the trench as can be clearly seen from the picture and the cross-section provided on the right panel of figure 4.

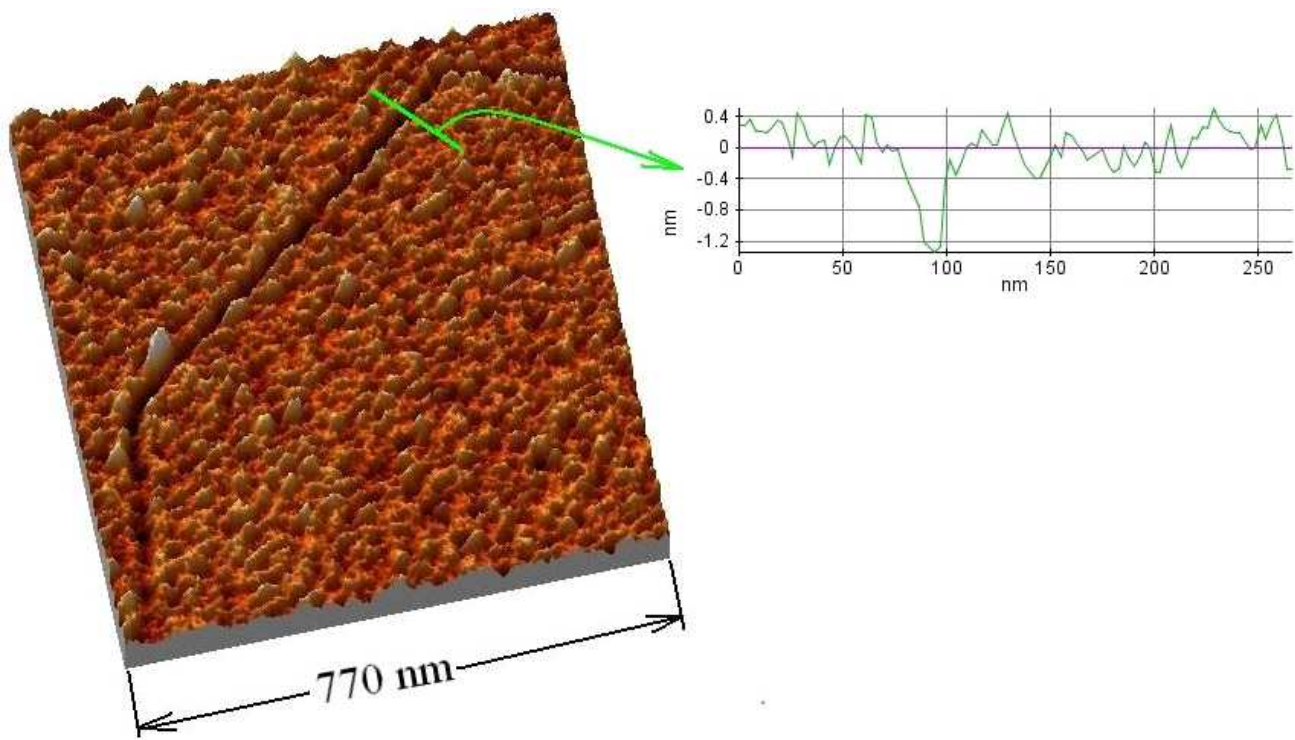

**Figure 4.** "Floating Tip Nanolithography" on a gold film (15 nm of Au on a Si wafer with 2 nm of Cr buffer layer). Writing speed, laser power and TSG are the same as those of figure 1.

**Experimental method and Floating Tip basics**
Our set-up is based on a commercial AFM XE-120 (Park Systems Corp.), custom designed for optical access and ASNOM applications. In the Z (vertical) direction, the height of the tip above the sample is controlled by a stacked piezoelectric actuator, physically separated from the X-Y (horizontal) scanner on which the sample is mounted. The system provides external remote access to practically all operational parameters. To monitor and control the TSG, we introduced very low amplitude (<1 nm) oscillations to the cantilever at a frequency far from its mechanical resonance, and used the measured response of the sensor for our feedback system. Figure 5 depicts the typical behavior of an Olympus AC160TS cantilever (spring constant 42 N/m, resonance frequency 342 kHz) near the surface. The black curve is the familiar dependence of the cantilever deflection on the distance, while the blue and red curves give the response (phase and amplitude respectively) of the cantilever to the oscillatory voltage (at frequency 75 KHz) driving the small oscillations as measured by the reflection of a probe beam from the back of cantilever into the Quadrant Position Sensitive Photo Detector (QPSPD). The cantilever mean deflection is measured by the dc component of the QPSPD response, whereas the amplitude and phase of the oscillations are extracted from the ac term.

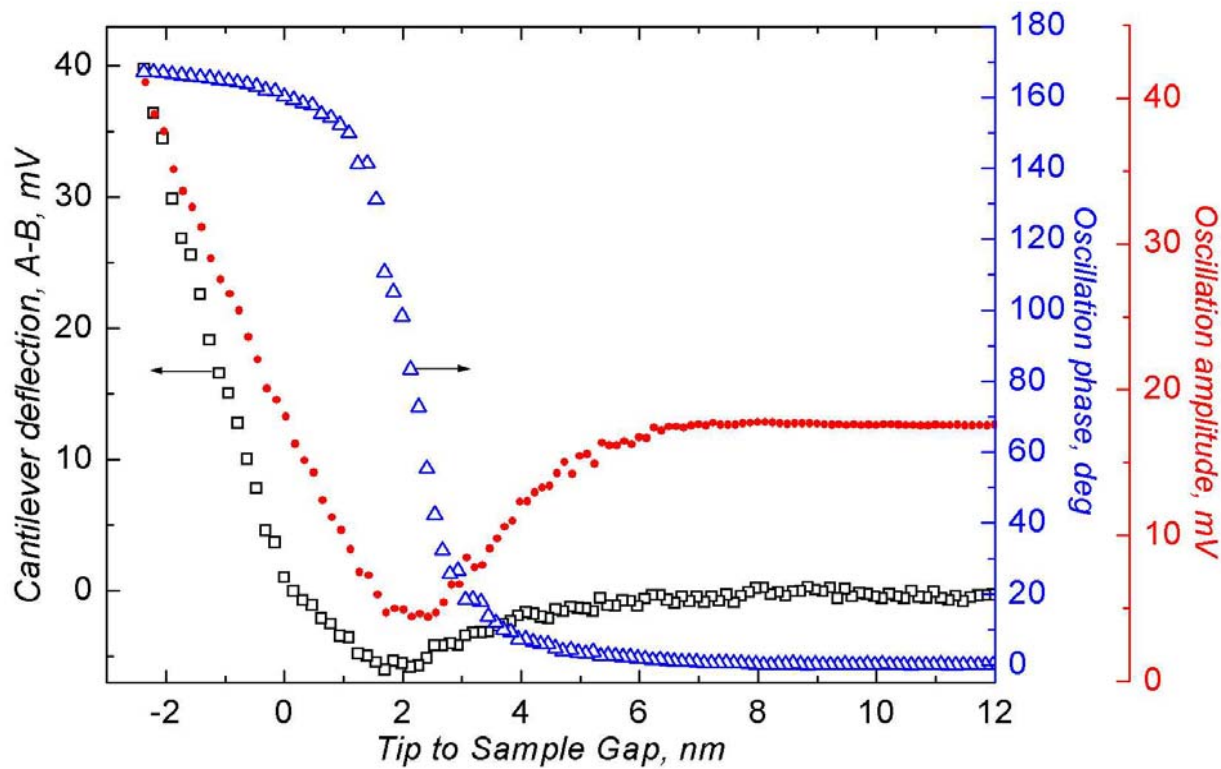

**Figure 5.** QPSPD measure of the cantilever deflection (black), phase (blue) and amplitude (red) response to small oscillation driving (<1 nm at 75 kHz) as a function of the TSG for a stiff cantilever (42 N/m).

While lowering the tip towards the sample, at a TSG of about 5 – 7 nm, the tip begins to feel the surface forces. As the tip approaches closer, the amplitude of the oscillations drops and a phase shift gradually builds up to 180 degrees which we interpret as a "hard contact". Intuitively, this change of phase is quite understandable: as the tip sets closer and closer to the surface, the pivot of cantilever motion shifts from one of its end to the other, providing a reversal of the phase. The curves in figure 5 are quite typical, even though the exact shape may depend slightly on the specific experimental conditions, such as humidity, sample material, position of the laser spot on the cantilever, etc. As can be seen in figure 5, the phase of the oscillations presents a single-valued function which is very suitable for TSG control in the range of 1 to 4 nm. To realize this control, we built a feedback circuit comprising of the lock-in detected signal from the QPSPD and the voltage fed to the Z-scanner.

The independent calibration of the relative displacement of the Z-scanner is straightforward. The calibration of the phase lag to an accuracy of ~1 nm in terms of the actual TSG, and in particular the determination of the TSG=0 point are much less trivial, but were possible by means of a procedure which is based on the scattering of the evanescent field at the interface of two media. The details of the experimental system, as well as a description of the calibration procedure, are given in Appendix A.

We use an ultrafast femtosecond laser (Micra, Coherent Inc.) providing pulses of 800 nm central wavelength, ~20 femtosecond duration at average power of up to 500 mW at a repetition rate of 80 MHz. Whereas for the hot tip effects we could have used a CW laser, the choice of the femtosecond laser stems from the wish to have high peak intensities so that we can observe nonlinear effects and electromagnetic field enhancement by the sharp tip. Finite element analysis of the tip temperature rise under irradiation, performed with parameters similar to our experimental conditions, provided information about the temperature rise, as well as the approach to thermal equilibrium. It was directly confirmed that the 'quasi-continuous' operation at a high repetition rate of laser pulses allows the tip to reach a thermal equilibrium, enabling us to calibrate out any miscellaneous effects such as cantilever bending and tip elongation, that may have caused ambiguity in earlier works. Last technical point: because the cantilever oscillates at a frequency that is very far from mechanical resonances, the heating doesn't affect the working curve 'phase vs. TSG' of the AFM.

**Conclusion**

The advantages of noncontact Floating Tip Nanolithography were demonstrated and discussed. We showed nano-writing and surface modification on polymer and metal films with spatial resolution of about 20 nm width and a few nm depth without direct contact with the sample surface. The new method enables reproducible, continuous tip surface interactions at a gap of a few nanometers, and can be used for contactless material processing. For the purpose of this controlled noncontact processing, we had to develop a new mode of operation for Atomic Force Microscopes which enables scanning a tip at a predetermined low height above the sample without touching the surface. Our novel approach to the control of the tip-sample gap is based on very small amplitude (<1 nm) forced oscillations of the cantilever at a frequency far from its main mechanical resonance. When the AFM tip is illuminated by a femtosecond laser, surface modification occurs based on either one of two mechanisms: a hot tip interaction with a surface, leading to the melting/evaporation of the material, or electromagnetic field enhancement under the tip triggering the material ablation. Future applications to other metallic surfaces, as well as to the initiation of surface chemical molecular modifications will be discussed in forthcoming publications.

**Acknowledgment**

We thank M. Karpovski for his invaluable help in the preparation of the gold films. This work was supported in part by the Nancy and Steve Grand Center for Sensors and Security, and by the James Franck program.

**Appendix A. Experimental set-up and Tip-Sample Gap calibration**

The sceleton scheme of the experimental set-up for the Floating Tip Nanolithography is shown in figure A1, and its variant for the Tip-Sample Gap (TSG) measurement and calibration – in figure A2.

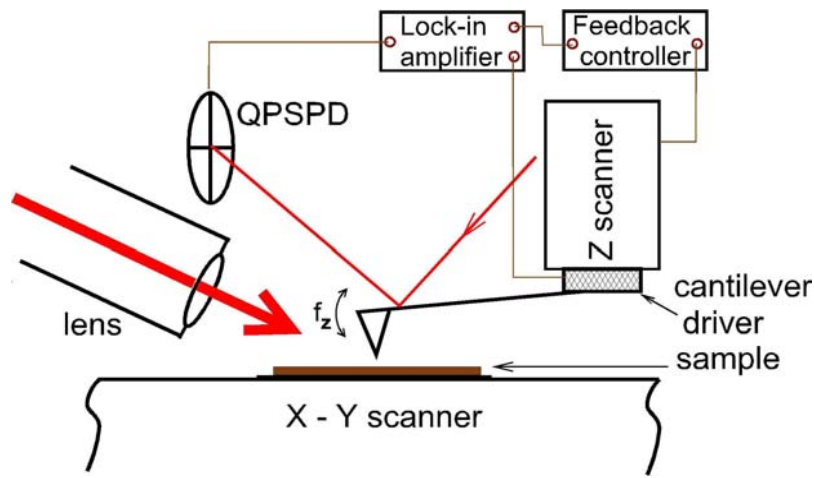

**Figure A1**. Experimental set-up.

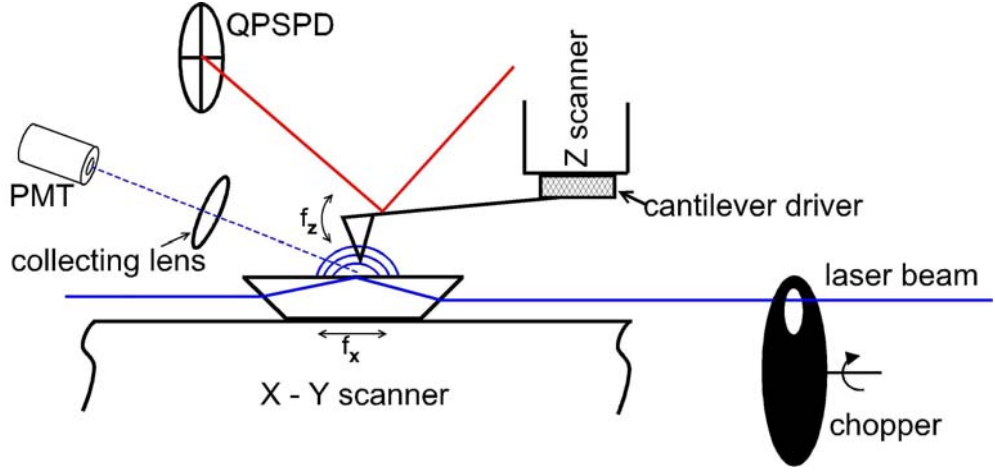

**Figure A2**. Set-up variation for the tip-sample gap calibration.

The independent variable in our experiments was the vertical travel of the Z-scanner, controlled with ~1 nm accuracy by the voltage on its piezo-ceramic body. As a measurable quantity uniquely dependent on the TSG, we chose the far field intensity of the tip-scattered evanescent field over a dove prism. This arrangement resembles common configurations involving total internal reflection, see e.g.[22]. The 5 mm glass-BK7 prism was mounted on the X-Y scanner of the AFM, with its longest face up, towards the Z-scanner with the probe. The 532 nm laser beam was directed horizontally into the $45^0$ side face of the prism. The laser was modulated with a mechanical chopper, and the far field light scattered by the tip dipped in the evanescent field over the dove prism surface was measured with a photo-multiplier and lock-in amplifier. The scattered light intensity is found to decrease exponentially with the tip height according to the Fresnel evanescent wave formula

$I = I_0 \times \exp(-2z/d_p)$

with the characteristic decay parameter (penetration depth)

$d_p/2 = \lambda / (2 \times 2\pi \times (n_2^2 \sin^2\varphi - n_1^2)^{1/2}) \approx 40$ nm

in good accordance with the known values of the refractive indices $n_1=1$, $n_2=1.52$ and incidence angle $\varphi = 72^0$. Special care was taken of the prism surface cleanliness and optical elements adjustment, which led to very effective background light suppression and very good signal to noise ratio (see figure A3).

The scattered light provided an independent, accurate measure of the TSG, which in turn was used to calibrate all other measurements carried out through the internal monitor laser reflected from the cantilever into the Quadrant Position Sensitive Photo Detector (QPSPD).

The 'standard' $(A-B)_{dc}$ signal from the QPSPD was monitored, providing information on the mean cantilever deflection. Using the piezo-electric driver, we forced small cantilever oscillations along the vertical axis at a frequency of a few kHz (far from any mechanical resonances of the cantilever) and amplitude of less than 1 nm. We tracked (lock-in detection) the amplitude and phase of the corresponding QPSPD signal typically referred to as $(A-B)_{ac}$. In addition, we measured the lateral shear force by shaking the sample holder along one horizontal axes at a low frequency (41 Hz) and amplitude of less than 1 nm, and monitored (lock-in detection) the response of the QPSPD corresponding to horizontal motion, typically referred to as $(C-D)_{ac}$. The above mentioned measurements, namely the $(A-B)_{dc}$, $(A-B)_{ac}$ (amplitude and phase), $(C-D)_{ac}$ (amplitude), and the far field light intensity were all taken simultaneously in each experimental run.

The operational procedure included the tip-sample contact approach followed by slow lift and then lowering back of the probe. Figure A3 presents the ascending part of a typical experimental data set obtained with a stiff cantilever (Olympus AC160TS, 42 N/m). The tip height reference (zero) point on the left panel of figure A3 is arbitrary, while the shift values along abscissa are derived from both the Z-scanner calibrated control voltage and internal Z-detector readings, and are accurate to better than 1 nm.

The tip-scattered light intensity followed the exponential law very well over the entire range of tip heights, with the exception of very close proximity to the surface. Within distances of a few nanometers, one should expect manifestation of various interactions between the tip and the surface, including Van der Waals and electrostatic forces and, most importantly, the capillary effect of the thin water layer at ambient environment. At these distances the Z-scanner and Z-detector readings cannot be trusted as real indicators of tip position, and one has to use an independent measurement of the TSG. In the right panel of figure A3, the abscissa values are corrected using the scattered evanescent field intensity curve in the vicinity of the gap opening. Here, in the zone of special interest when the surface "releases" the tip, cantilever deflection, lateral cantilever oscillations and the phase of vertical low-amplitude forced oscillations showed unambiguous characteristic behavior. The steep change of the lateral amplitude which measures the lateral force (red triangles) indicates the moment of loss of "direct mechanical contact". The phase of vertical cantilever oscillation (black circles) reaches 180 degrees at the same time, *and we define this TSG as zero*. The range where the phase changes steeply overlaps the region of most interest 2 - 4 nanometers, and is therefore suitable as the control parameter. (For comparison: the ordinary non-contact mode of an AFM operates on the mechanical resonance of the cantilever. The amplitude of the cantilever oscillation is several tens of nanometers. When the tip starts to feel the surface at a distance of hundreds of nanometers (!), the amplitude reduces and the phase is changed, and therefore these large, resonant oscillations cannot be used for sensitive control at very short distances of a few nanometers).

As detailed in the text, the conventional method of a 'hard' approach followed by a lift cannot work with a soft tip. In figure A4 the result for the soft cantilever (Olympus AC240TS, spring constant 2 N/m, resonance frequency ~70 kHz) is presented. Note that as before, the "zero" height is defined by the point where the exponentially decaying curve of the scattered evanescent field meets the baseline level of a tip in contact. As a guide, let us follow the approaching tip. Upon descent (upper blue curve) the tip is 'snapped' by the surface (point 'a') at a height of ~18 nm, and remains stuck at the surface for the remainder of the descent. At this point, the intensity of scattered light (magenta squares) jumps to the contact level. Upon ascent, the large hysteresis is clearly observed – tip-surface contact is maintained for much longer, and the tip actually leaves the surface (point 'b') at a distance of ~85 nm, as also evidenced by the jump of the scattered light intensity. The very large hysteresis is characteristic for this kind of probes at ambient environment.

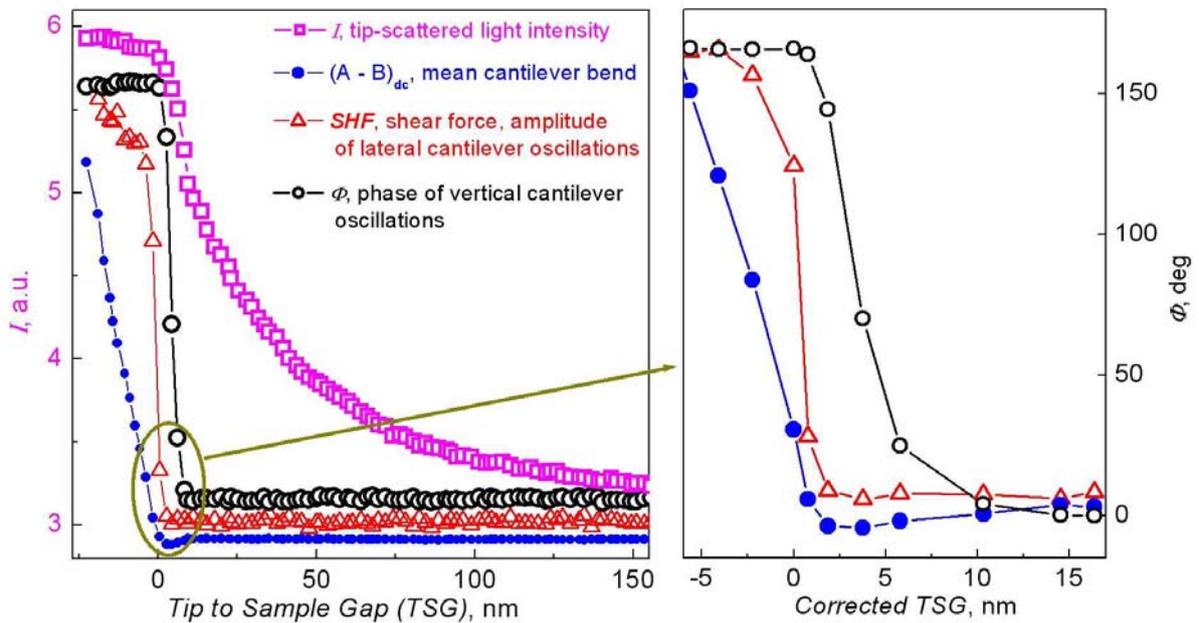

**Figure A3.** Tip height dependence of all measured variables at tip movement off the surface. The various lines are defined within the figure. The abscissa values in the right panel are corrected using the TSG dependence of scattered evanescent field intensity. All data except phase of vertical cantilever oscillations are in arbitrary units.

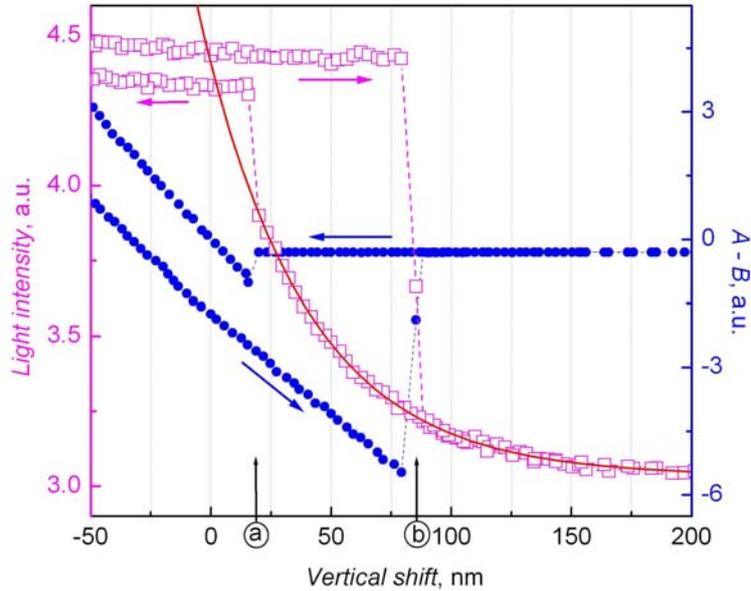

**Figure A4.** Vertical probe shift dependence of cantilever bend (blue) and scattered evanescent field intensity (magenta) for "soft" cantilever. Solid red line presents the exponential fit to the light intensity decay.

**Appendix B. Estimation of tip temperature**

Finite element analysis has been used to estimate the temperature rise and distribution in a laser heated tip. In the model, the tip is represented as a pyramid at the end of the cantilever while the other end of the cantilever was considered connected to a heat reservoir at room temperature (figure B1). Standard values of silicon thermo-physical and optical properties were used[23]. With the actual laser repetition rate of 80 MHz, steady state is reached after about 40,000 laser pulses, 0.5 msec (figure B2).

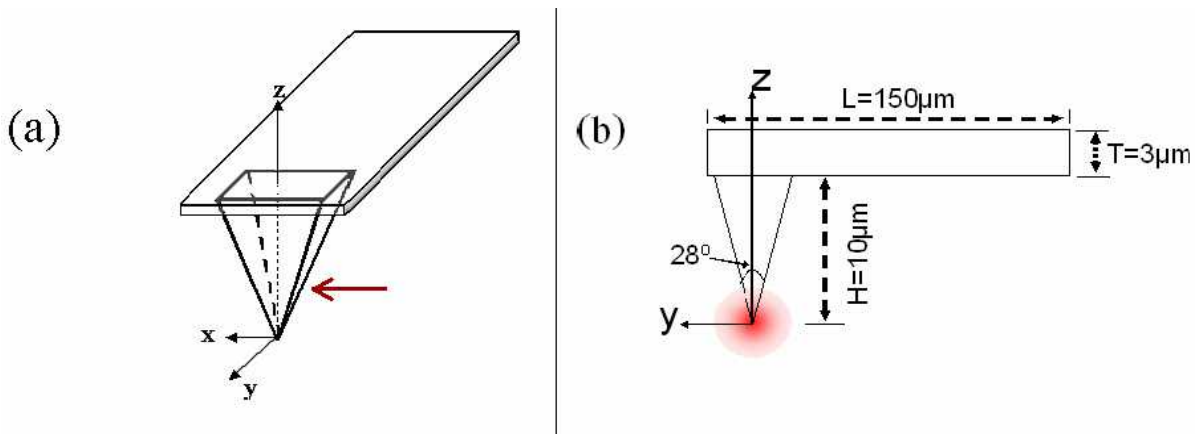

**Figure B1.** The geometrical arrangement (a) and probe dimensions (b) used in the calculation. Numbers are similar to the Olympus AC160TS probe employed in our experiments.

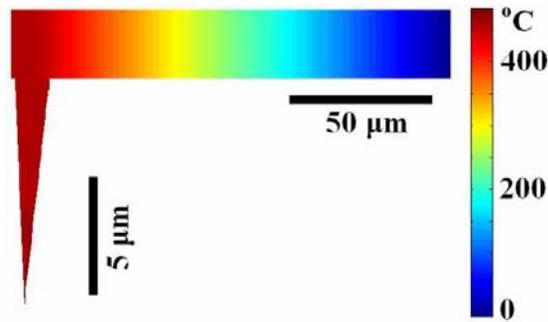

**Figure B2.** Steady state temperature distribution along the silicon tip and cantilever. A laser beam with energy fluence of 15 mJ/cm2/pulse is focused to a 3 micron spot on the apex of the tip. Note the different linear scales in vertical and horizontal directions.

The dependence of the temperature on the applied laser fluence is shown in figure B3. The linear absorption coefficient of silicon at 800 nm (1021 cm$^{-1}$) gives rise to approximately 1% absorption, and with the two photon absorption rate of 55x10$^{-9}$ cm/W, a weak deviation from linearity is observed at the highest peak powers.

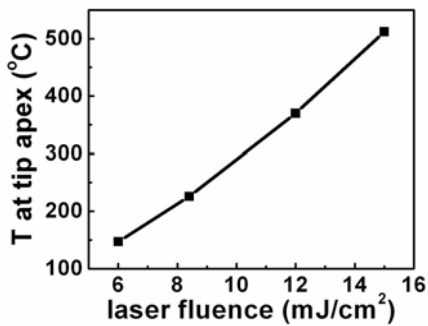

**Figure B3.** Steady state tip temperature as a function of laser fluence. The deviation from a linear dependence at higher powers is due to two photon absorption.